# Effect of the nitrogen-argon gas mixtures on the superconductivity properties of reactively sputtered molybdenum nitride thin films

N. Haberkorn,[1,2] S. Bengio,[1] S. Suárez,[1,2] P. D. Pérez,[1] M. Sirena,[1,2] J. Guimpel.[1,2]

[1]*Comisión Nacional de Energía Atómica and Consejo Nacional de Investigaciones Científicas y Técnicas, Centro Atómico Bariloche, Av. Bustillo 9500, 8400 San Carlos de Bariloche, Argentina.*

[2] *Instituto Balseiro, Universidad Nacional de Cuyo and Comisión Nacional de Energía Atómica, Av. Bustillo 9500, 8400 San Carlos de Bariloche, Argentina.*

## Abstract

We report on the superconducting properties of molybdenum nitride thin films grown by reactive DC sputtering at room temperature with a $N_2$:Ar mixture. Thin films grown using 5 % $N_2$ concentration display $T_c$ = 8 K, which is gradually reduced and abruptly disappears for 40 % $N_2$ concentration. This suppression can be associated with changes in the nitrogen stoichiometry from $Mo_2N$ to MoN. Our results provide an effective and simple path to prepare $Mo_2N_x$ thin films with tunable $T_c$, which is relevant for the investigation of the fundamental properties and for technological applications.

*e-mail*: *nhaberk@cab.cnea.gov.ar*; Tel:+540294 4445147-FAX:+540294 4445299

## 1. *Introduction*

Transition–metal nitrides (TMN) display a wide range of electronic and mechanical properties which are promising for technological applications. Superconducting TMN are potential candidates in a wide range of cryogenic devices like tunnel junctions [1] and electromagnetic radiation detectors [2]. The Mo nitrides present several superconducting crystalline phases: $\gamma$-$Mo_2N$ (cubic) with $T_c \sim$ 5 K [3], $\beta$-$Mo_2N$ (tetragonal) with $T_c \sim$ 5 K [4] and $\delta$-MoN (hexagonal) with $T_c \sim$ 12 K [5]. Different methods have been used in the growth of Mo nitride thin films, such as reactive sputtering [6,7], pulsed laser deposition [8], thermal nitration [9] and chemical routes [10]. A distinctive feature of $\gamma$-$Mo_2N$ thin films is the influence of the disorder on $T_c$, which ranges from 4.5 K  to around 8 K for epitaxial and polycristalline thin films, respectively [11,12].

In this letter, we show that the $T_c$ in $\gamma$-$Mo_2N_x$ thin films (grown by DC sputtering at room temperature) can be tuned by modifying the $N_2$:Ar mixture used during the sputtering process. $T_c$ in thin films can be modified from 8 K to temperatures below 3 K by increasing the $N_2$ partial pressure in the $N_2$:Ar mixture from 5% to 40% of the total pressure. This modification can be associated with changes in the nitrogen stoichiometry from $Mo_2N$ to MoN.

## 2. *Material and methods*

Mo$_2$N$_x$ films were deposited by DC reactive magnetron sputtering, starting from a pure Mo target in a N$_2$:Ar gas mixture with N$_2$ partial pressure going from 5% to 70% of the mixture's 5 mTorr pressure. Films were grown on top of an 8 nm thick AlN buffer layer (grown with 20% N$_2$ partial pressure), introduced to avoid any chemical reaction between the Mo and the native SiO$_2$ layer of the Si wafers. During deposition the target to substrate distance was ~ 5.5 cm. The target power was 100 W (AlN) and 50 W (Mo-N). Ultra-high purity Ar (99.999%) and N$_2$ (99.999%) were used as gas sources. The residual pressure of the chamber was less than $10^{-6}$ Torr. No intentional heating of the substrate was used. Wherever used, the notation [MoNY%] indicates a Mo-N film grown with Y% N$_2$ partial pressure.

X-ray (XRD) difracction data were obtained using a Panalytical Empyrean equipment. The chemical composition and thicknesses of the films were analyzed by Rutherford Backscattering Spectroscopy (RBS) with a TANDEM accelerator using a 2 MeV $^4$He$^{2+}$ ion beam. RBS spectrums were simulated with the SIMNRA code. Surface composition analysis was performed by means of X-ray photoelectron spectroscopy (XPS) using a standard Al/Mg twin-anode, X-ray gun and a hemispherical electrostatic electron energy analyzer. The electrical transport measurements were performed using the standard four point configuration.

## 3. *Results and discussion*

XRD for thin films grown with N$_2$/Ar mixtures ≤ 40 % show the 200 reflection corresponding to the cubic γ-Mo$_2$N structure (see Fig. 1*a*). The peak is systematically sifted to smaller angles for higher N$_2$ partial pressure, which indicates a larger lattice parameter (see Table 1). The microstructure of the films displays columnar nanometric grains (diameter smaller than 10 nm) textured along the (100) axis [13]. No X-ray reflections were identified for N$_2$/Ar mixtures > 40%, that could indicate a change in the structure of the films. For Mo-N films grown with different N$_2$ partial pressures, the electrical resistivity increases with T between 10 K and 300 K. The residual resistance ratio ($RRR = R^{300K}/R^{10K}$, with $R$ the sample resistance) is in the range 0.96-0.6, systematically decreasing for higher N$_2$ partial pressures (see Table I). This indicates a very short mean free path $l$ due to structural disorder (grain boundaries and chemical impurities). Figure 1*b* shows the dependence of $T_c$ with the N$_2$ partial pressure. The inset in Fig. 1*b* shows the temperature dependence of the normalized resistance at $T < 10$ K for the superconducting films. The results show that for [MoN5%], $T_c$ is close to 8 K and it systematically decreases until disappearing for [MoN40%], The superconducting transition is sharp for all samples. Considering that all the analyzed films are thicker than 40 nm, no contribution of their dimension on the $T_c$ is expected [13].

In order to understand the influence of the N$_2$:Ar mix on the stoichiometry of the films, we have analyzed the chemical composition by RBS (see inset Fig. 1*c*). The drop in $T_c$ can be correlated with changes in the Mo/N chemical composition (see Fig. 1c). The RBS results show that for [MoN5%] the stoichiometry is close to Mo$_2$N and it systematically decreases to MoN at [MoN40%]. MoN stoichiometry is expected for δ-MoN [10].

The absence of $T_c$ for $N_2/Ar > 40\%$ could be attributed to an increment in the disorder and no crystallization of the hexagonal δ-MoN. It is important to note that epitaxial γ-Mo$_2$N films with $RRR \approx 50$ display $T_c = 4.5$ K [11] and polycrystalline γ-Mo$_2$N films with $RRR < 1$ display $T_c = 8$ K [11]. In addition, the $T_c$ in [MoN5%] annealed up to 973 K decreases from 8 K to ≈ 5 K, which is close to the value reported for epitaxial films [11]. This indicates that for Mo-N thin films grown at room temperature the stoichiometry is a relevant parameter for $T_c$. On the other hand, increasing the $N_2$ partial pressure affects the sputtering rate, which decreases from ≈ 22 nm / min for [MoN5%] to ≈ 10 nm/min for [MoN70%] (see Fig. 1*d*).

XPS measurements were performed to obtain information of the chemical composition of the Mo-N films and the oxidation state of the Mo. The photoelectron peaks Mo3d, O1s, C1s and N1s were measured in detail. An overlapping of the N1s and Mo3p peaks was observed. The Mo3d binding energy region for each MoN film is shown in Figs. 2*a-d*. The spectra were fitted using a Voight function for each peak plus a Shirley-type background. The total fitted intensities along with the experimental ones are shown in each spectrum. In the [MoN10%] and [MoN20%] Mo3d spectra, up to three components were identified: a low binding energy component at 228.5 eV (which could be ascribed to Mo$^{\delta+}$ ($2 \leq \delta < 4$) associated with the compound Mo$_2$N [14]), an intermediate component at ≈ 230 eV (which could be attributed to Mo$^{4+}$ associated to MoO$_2$ impurities [15]), and a high binding energy component at 232.7 eV (ascribed to the presence of Mo$^{6+}$ due to the formation of MoO$_3$ in the surface [16,15]). The doublet associated to the presence of MoO$_3$ can be completely removed after a sputtering process, which indicates its surface nature [13]. The Mo3d spectra at [MoN10%] and [MoN20%] are dominated by Mo$^{\delta+}$. In the [MoN35%] and [MoN50%], Mo3d spectra display a new component at 229.1eV, which can be ascribed to MoN [17]. This new component shifts the envelope of the Mo3d$_{5/2}$ peak to higher binding energy, as observed by Wang *et al.* [14].

Figure 3 shows $H_{c2}(T)$ with the magnetic field perpendicular to the surface for [MoN5%], [MoN10%], [MoN20%] and [MoN30%]. Inset Fig. 3 shows typical resistivity vs. temperature curves for different applied magnetic fields ($H$) in [MoN30%]. The temperature dependence of $H_{c2}$ can be analyzed by the Werthamer-Helfand-Hohenberg (WHH) model developed for dirty one-band superconductors [18], which predicts

$$ln\frac{1}{t} = \sum_{v=-\infty}^{\infty} \left( \frac{1}{|2v+1|} - \left[ |2v+1| + \frac{\hbar}{t} + \frac{(\alpha\hbar/t)^2}{|2v+1| + (\hbar+\lambda_{so})/t} \right]^{-1} \right) \text{ [eq. 1]},$$

where $t = T/T_c$, $\hbar = (4/\pi^2)\left(H_{c2}(T)/|dH_{c2}/dT|_{T_c}\right)$, $\lambda_{so}$ is the spin-orbit scattering constant, and $\alpha$ is the Maki parameter which quantifies the weakening influence of the Pauli electron spin paramagnetism on the superconducting state. When $\lambda_{so} = 0$, $H_{c2}(0)$ obtained from the WHH formula satisfies the relation $H_{c2}(0) = \frac{H_{c2}^{orb}(0)}{\sqrt{1+\alpha^2}}$, which is originally derived by K. Maki [19]. All the analyzed $H_{c2}(T)$ curves followed the WHH model with $\alpha = 0$ and $\lambda_{so} = 0$. For $\alpha = 0$, $H_{c2}(T)$ is given as the pure "orbital field limit", $H_{orb}(T)$, due to the supercurrents circulating around the vortex cores. The results show that, independently of $T_c$, all the films present $H_{c2}(0)$ around 12 T (see Fig. 3 and Table I). This value is close to the one expected from the Pauli limit $H_p \approx 1.84\ T_c$ [20] and corresponds to a coherence length of $\xi(0) \approx 5.2$ nm.

## 4. Conclusions

In summary, nanocrystalline superconducting $Mo_2N_x$ thin films with $RRR < 1$ have been successively grown at room temperature by reactive DC magnetron sputtering. The $T_c$ of the films can be tuned from 8 K to below 3 K by modifying the reactive $N_2$:Ar composition. The drop in $T_c$ can be attributed to changes in the Mo/N stoichiometry in the films. Thin films with $T_c > 6$ K display $H_{c2}(0) \approx 12$ T and $\xi(0) = 5.2$ nm.

## Acknowledgments

This work was partially supported by the ANPCYT (PICT 2015-2171), U. N. de Cuyo 06/C505 and CONICET PIP 2015-0100575CO.

Figure 1. Dependence with the $N_2$:Ar reactive gas composition for films grown in a lapse of 5 minutes of : *a) XRD; b)* $T_c$*; c)* Atomic Mo/N ratio; *d)* Sputtering growth rate [nm/min]. Inset *b)* Normalized resistance $R/R^{10K}$ for the superconducting films at $T < 10$ K. The arrow indicates the change in the $N_2$:Ar mixture from 5% to 35%. Inset *c)* RBS spectra for [MoN50%].

Figure 2. XPS Mo3d spectra of: *a)* [MoN10%]; *b)* [MoN20%]; *c)* [MoN35%]; and *d)* [MoN50%].

Figure 3. Temperature dependence of the upper critical field ($H_{c2}$) for [MoN5%], [MoN10%], [MoN20%], and [MoN30%]. Inset: Resistivity vs. Temperature dependence for different $H$ in [MoN30%].

Table I. List of samples and related properties.

Table 1.

| Sample | $a$ [nm] | $RRR$ | $T_c$[K] | $H_{c2}(0)$ [T] | $-\delta H_{c2}/\delta t\mid_{T_c}$ |
|---|---|---|---|---|---|
| [MoN5%] | - | 0.96 | 7.95 (0.05) | 12.4 (0.1) | 17.8 (0.1) |
| [MoN10%] | 0.4202 (0.0005) | 0.95 | 7.20 (0.05) | 11.8 (0.1) | 17 (0.2) |
| [MoN20%] | 0.4230 (0.0005) | 0.93 | 6.6 (0.1) | 11.8 (0.1) | 17 (0.2) |
| [MoN30%] | 0.4280 (0.0005) | 0.88 | 5.80 (0.05) | 11.8 (0.1) | 17 (0.2) |
| [MoN35%] | 0.4300 (0.0005) | 0.84 | 4.20 (0.05) | - | - |
| [MoN40%] | 0.4330 (0.0005) | 0.81 | < 3 K | - | - |
| [MoN50%] | No reflections | 0.8 | - | - | - |
| [MoN60%] | No reflections | 0.75 | - | - | - |
| [MoN70%] | No reflections | 0.6 | - | - | - |

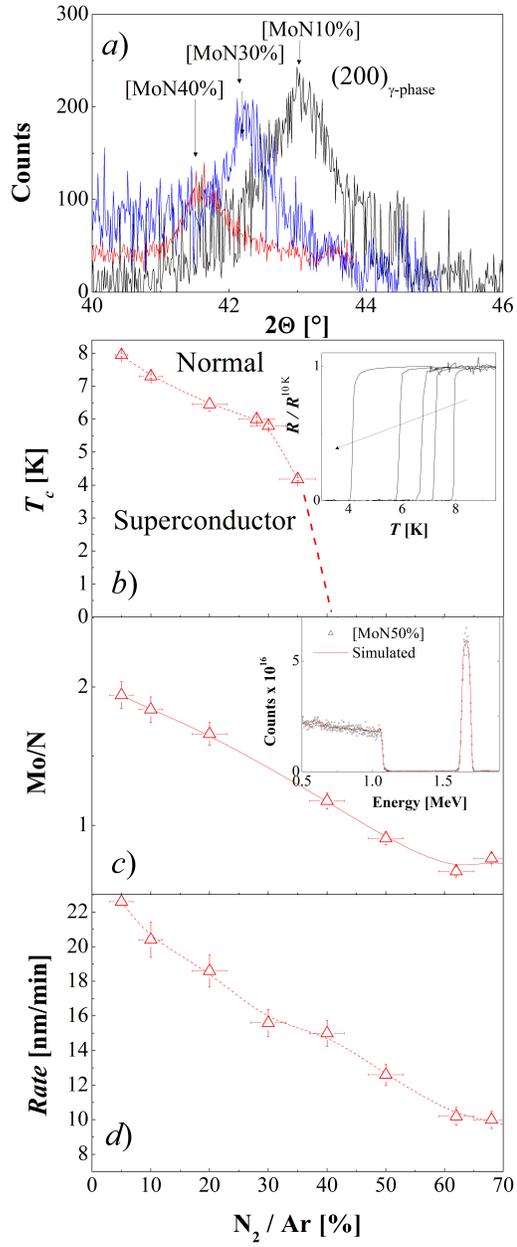

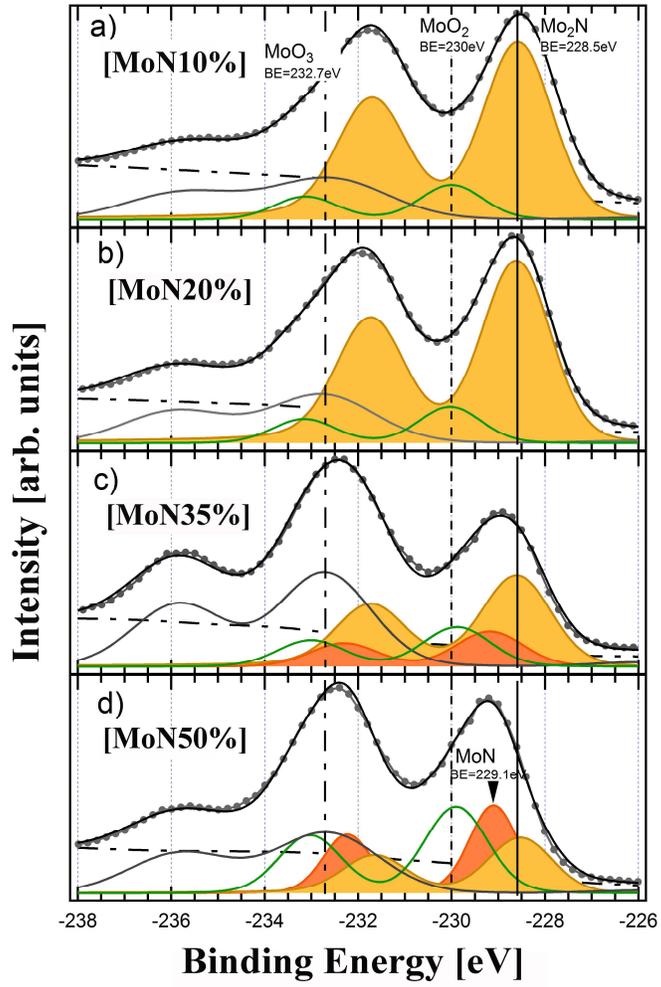

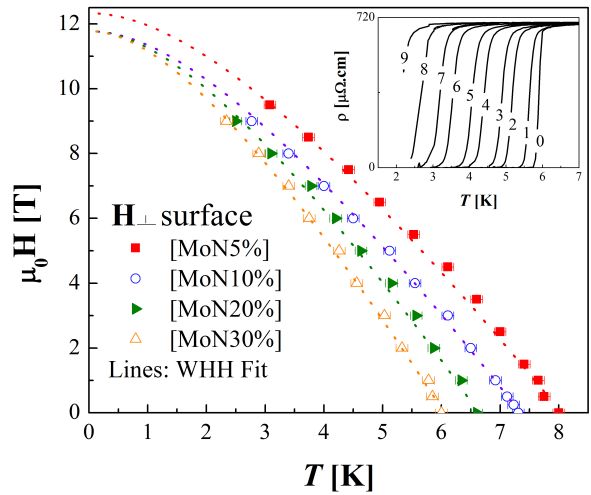

[1] Zhen Wang *et al.*, Appl. Phys. Lett.**75** (1999) 701-703.

[2] Chandra M Natarajan *et al.*, Supercond. Sci. Technol. **25** (2012) 063001- 063001(16).

[3] B. T. Matthias, J. K.Hulm, Phys. Rev. **87** (1952) 799-806.

[4] Kei Inumaru *et al.*, Chem. Mater. **17** (2005) 5935-5940.

[5] Shanmin Wang *et al.*, Scient.Rep **5** (2015) 13733-13733(8).

[6] Y. H. Shi *et al.*, Phys. Rev B **38** (1988) 4488-4491.

[7] H. Ihara *et al.*, Phys. Rev. B **32** (1985) 1816-1817.

[8] Kei Inumaru *et al.*, Phys. Rev. B **73** (2006) 52504-52504(4).

[9] D K Christen *et al.*, IEEE Trans. Magn., **23** (1987) 1014-1018.

[10] Y. Y. Zhang *et al.*, *J. Am. Chem. Soc.* **133** (2011) 20735-20737.

[11] Hongmei Luo *et al.* J. Phys. Chem. C **115** (2011) 17880-17883.

[12] R. Baskaran *et al.*, J. Phys. D **49** (2016) 205304-205307.

[13] N. Haberkorn *et al.* (unpublished).

[14] Y. M. Wang, R. Y. Lin, *Mater. Sci. Eng.* B **112** (2004) 42-49.

[15] Z. B. Zhaobin Wei, P. Grange, B. Delmon. Appl. Surf. Scie **135** (1998) 107-114.

[16] G.-T Kim *et al.*, Appl. Surf Scie **152** (1999) 35-43.

[17] Kejun Zhang *et al.*, Appl. Mat. Interfaces **5** (2013) 3677-3682.

[18] N. R. Werthamer, E. Helfand and P. C. Hohenberg, Phys. Rev. **147** (1966) 295-302.

[19] K. Maki, Phys. Rev. **148** (1966) 362-369.

[20] M. Tinkham, Introduction to Superconductivity 2nd edn (McGraw-Hill, 1996).